\PassOptionsToPackage{hyphens}{url}
\documentclass{aa}
\usepackage{graphicx}
\usepackage{txfonts}
\usepackage{amsmath,amssymb}
\usepackage{lmodern}
\usepackage{iftex}
\usepackage{placeins}
\ifPDFTeX
  \usepackage[T1]{fontenc}
  \usepackage[utf8]{inputenc}
  \usepackage{textcomp} 
\else 
  \usepackage{unicode-math}
  \defaultfontfeatures{Scale=MatchLowercase}
  \defaultfontfeatures[\rmfamily]{Ligatures=TeX,Scale=1}
\fi
\IfFileExists{upquote.sty}{\usepackage{upquote}}{}
\IfFileExists{microtype.sty}{
  \usepackage[]{microtype}
  \UseMicrotypeSet[protrusion]{basicmath} 
}{}
\makeatletter
\renewcommand*\aa@pageof{, page \thepage{} of \pageref*{LastPage}}
\@ifundefined{KOMAClassName}{
  \IfFileExists{parskip.sty}{%
    \usepackage{parskip}
  }{
    \setlength{\parindent}{0pt}
    \setlength{\parskip}{6pt plus 2pt minus 1pt}}
}{
  \KOMAoptions{parskip=half}}
\makeatother
\usepackage{xcolor}
\IfFileExists{xurl.sty}{\usepackage{xurl}}{} 
\IfFileExists{bookmark.sty}{\usepackage{bookmark}}{\usepackage{hyperref}}
\hypersetup{
  hidelinks,
  pdfcreator={LaTeX via pandoc}}
\makeatletter
\def\maxwidth{\ifdim\Gin@nat@width>\linewidth\linewidth\else\Gin@nat@width\fi}
\def\maxheight{\ifdim\Gin@nat@height>\textheight\textheight\else\Gin@nat@height\fi}
\makeatother
\setkeys{Gin}{width=\maxwidth,height=\maxheight,keepaspectratio}
\makeatletter
\def\fps@figure{htbp}
\makeatother
\ifLuaTeX
  \usepackage{selnolig}  
\fi

\author{}
\date{}
\begin{document}

\title{On the origin of the type-III radiation observed near the Sun}

\author{F.S. Mozer\inst{1,2} \and
O. Agapitov\inst{1} \and
S.D. Bale\inst{1,2} \and
K. Goetz\inst{3} \and \\
V. Krasnoselskikh\inst{4} \and
M. Pulupa\inst{1} \and
K. Sauer\inst{5} \and
A. Voshchepynets\inst{6}}

\institute{
Space Sciences Laboratory, University of California, Berkeley,
California \\
\and
Physics Department, University of California, Berkeley, California \\
\and
University of Minnesota, Minneapolis, Minnesota \\
\and
LPC2E, CNRS-University of Orléans-CNES, 45071, Orléans, France \\
\and
Max-Planck Institute of Solar System Research, Göttingen, Germany \\ 
\and
Uzhhorod National University, Uzhhorod, Ukraine  \\
}
%
\abstract{}{ 
   
      To investigate processes associated with generation of type-III radiation.
    
}{ 

    Measure the amplitudes and phase velocities of Parker Solar Probe observed electric fields, magnetic fields, and plasma density fluctuations.
}{ 

  1.  There are slow electrostatic waves near the Langmuir frequency and at as many as six harmonics.  Their wave number is estimated to be 0.14 and $k \lambda_d$ = 0.4.  Even with a large uncertainty in this quantity (more than a factor of two) the phase velocity of the Langmuir wave was <10,000 km/sec. 
  
    2. There is an electromagnetic wave near the Langmuir frequency having a phase velocity less than 50,000 km/sec.
    
    3.  Whether there are electromagnetic waves at the harmonics of the Langmuir frequency cannot be determined because the amplitudes of their magnetic fields would be below the instrument threshold.
    
    4.  The less- than-one-millisecond amplitude variations typical of the Langmuir wave and its harmonics are artifacts resulting from addition of two waves, one of which has small frequency variations that arise from propagation through density irregularities.  
    
None of these results are expected in the conventional model of the three-wave interaction of two counter-streaming Langmuir waves that coalesce to produce the type-III wave.  They are consistent with a new model in which electrostatic antenna waves are produced at the harmonics by radiation of the Langmuir wave, after which at least the first harmonic wave evolved through density irregularities until its wave number decreased and it became the type-III electromagnetic radiation.}{}
\maketitle

\section{Introduction}

An important wave near the Sun is the Type-III radiation, which is an
electromagnetic wave whose frequency decreases with time and distance from the Sun. The theory of
type-III bursts was described by Ginsburg and Zhelezniakov {[}1958{]} as
a multi-step process in which Langmuir waves are produced by an electron
beam, after which some of these waves interact with density
irregularities to become backward waves that coalesce with the forward
waves in a three-wave process that produces the electromagnetic type-III
radiation. The theory has subsequently been discussed and refined by
many authors (e.g. Sturrock 1964; Zheleznyakov \& Zaitsev 1970; Smith
1970; Smith et al. 1976; Melrose 1980c; Goldman 1983; Dulk 1985; Melrose
1987{]}.
The first in-situ spectral observations of plasma waves in association
with type-III emissions were made by Gurnett and Anderson {[}1976{]}.
Their simultaneous observation of Langmuir waves, harmonic electric
field emissions, and electromagnetic waves was presented as
evidence of the direct conversion of pairs of Langmuir waves into
type-III waves, although the direct, in-situ, evidence of this
interaction was not and has not been obtained.

\section{Data}

On March 21, 2023, the Parker Solar Probe was located about 45 solar
radii from the Sun when it became embedded in the active type-III
emission region illustrated in Figure 1a. Note the colored lines under
the emission (at the plasma frequency) which indicate times when large
electric fields appeared. The type-III emission covered the frequency
range from about 200 to 400 kHz at the time of about 26 bursts of
high-rate data illustrated in Figure 1b. These bursts each consisted of
\textasciitilde15 milliseconds of electric and magnetic field
measurements at a data rate of about 1.9 million samples/second {[}Bale
et al, 2016{]}. That the emission range covered more than just twice the
plasma frequency has been noted earlier {[}Kellogg, 1980; Reiner and
MacDowell, 2019; Jebaraj et al, 2023{]} and this aspect of the emission
will be investigated below.
\begin{figure}[tb] 
\centering
\includegraphics{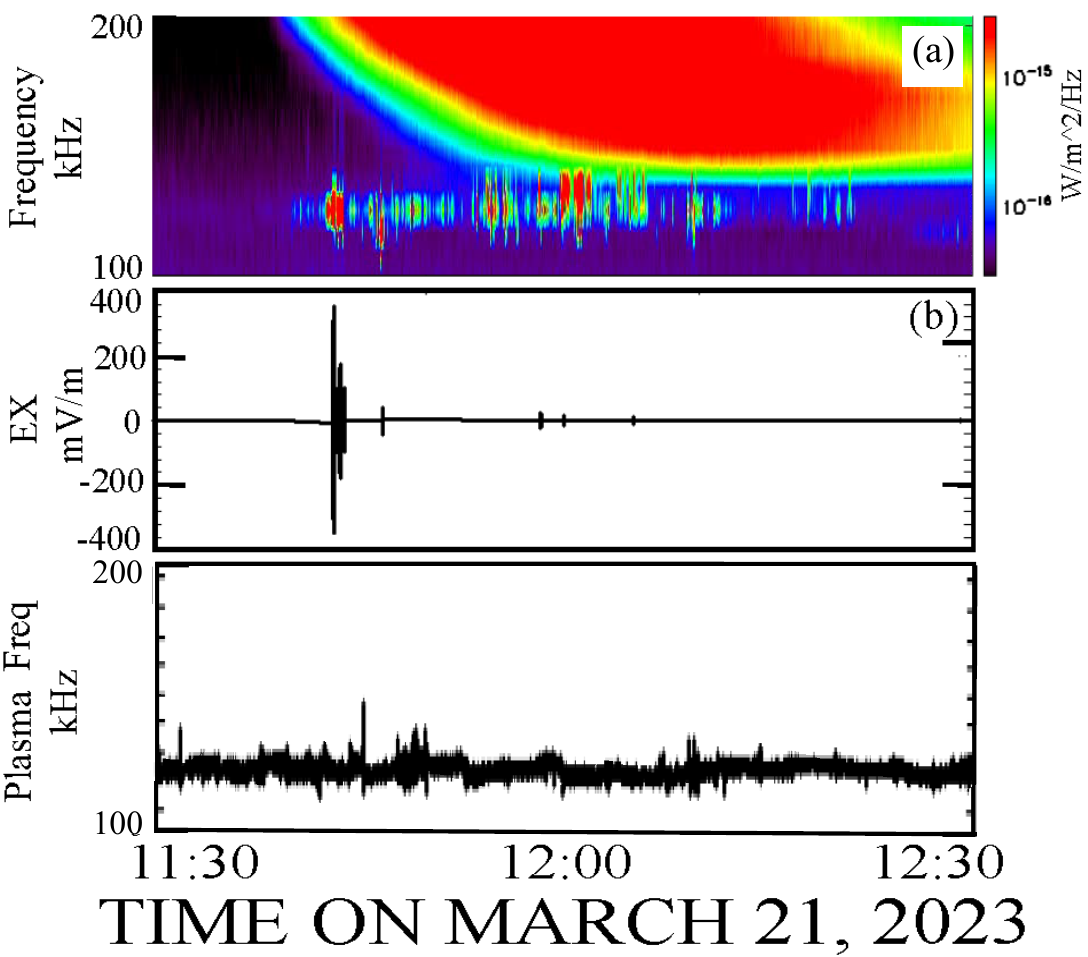}
\caption{
Overview of a type-III emission observed on March 21, 2023
(panel 1a) in which the spacecraft was embedded, showing bright lines
below the emission due to large electric fields. From the plasma
frequency of 125 kHz in Figure 1c, it is seen that these bright lines
were due to waves at or near the plasma frequency. Note that these waves
are present only part of the time. Figure 1b shows the times and
electric field amplitudes of 26 data bursts taken for duration's of about
15 milliseconds each at data rates in excess of one million
samples/second.
}
\end{figure}

The local plasma frequency, illustrated in Figure 1c, was determined as
$8980N^{0.5}$, where $N$ is the plasma density obtained from
a least-squares fit of the spacecraft potential to the square root of
the density obtained from the quasi-thermal noise measurement {[}Pedersen et al. 1984; Mozer
et al, 2022{]}. The relationship between the spacecraft potential and the plasma density results from the spacecraft potential being determined by equating the flux of photo-electrons that overcome the potential (to escape to infinity) with the incoming flux due to the random thermal electron current. Comparison of the frequency of Figure 1c with the vertical lines
beneath the type-III emission in Figure 1a shows that the waves observed
in the bursts were at or near the local plasma frequency.

To interpret these observations in more detail, eleven short bursts of data obtained during a 250 msec interval during the type-III emission are illustrated in Figures 2a through 2d. Panels 2a and 2b give the two
electric field components in the spacecraft X-Y plane, which is the
plane perpendicular to the Sun-satellite line.  Figure 2c plots the density fluctuations, $\Delta N / N$. Figure 2d presents the angle between the Langmuir $k$-vector and the
magnetic field in the X-Y plane. Because $B_z$ was almost
zero at this time, the error associated with ignoring the
out-of-the-X-Y-plane components in this measurement is small. It shows
that the waves were oblique to the magnetic field much of the time.  The largest event in Figure 2, near 11:43:15.460 had a maximum electric field of 300 mV/m and is discussed in the following sections. The small and medium size events near 11:43:15.450 and 11:43:15.580 are discussed in the appendix.

\begin{figure}[tb] 
\centering
\includegraphics[width=\linewidth]{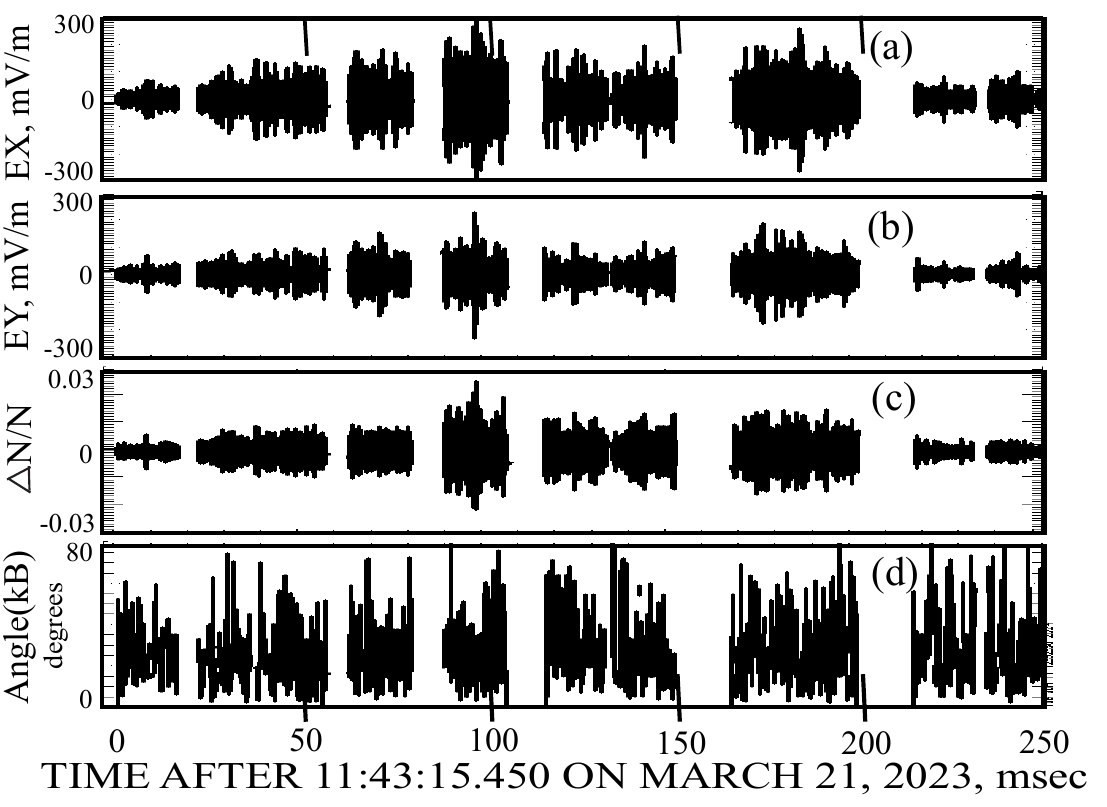}
\caption{
Eleven short bursts of data obtained during a 250 msec interval during the type-III emission. Figures 2a and 2b give the X and
Y components of the electric field in the spacecraft frame. Figure 2c gives the
density fluctuations that are expected to accompany  Langmuir waves.
Figure 2d presents the angle in the X-Y plane between the background
magnetic field and the electric field wave vector.  The largest event, near 11:43:15.460 is discussed in the main text and the small and medium size events near 11:43:15.450 and 11:43:15.580 are discussed in the appendix.
}
\end{figure}

In a Langmuir wave, the relationship between the density fluctuations and the electric potential is given as {[}Bellan, 2006{]}
\begin{equation}
\frac{\Delta N}{N} = \frac{e \phi}{m_e \omega_p^2 / k_p^2}
\end{equation}
where $\phi$ is the potential of the wave, $k_p$ is the wave number of 
the wave, and  $\omega_p = 2\pi\times 125$ kHz is the Langmuir wave frequency.
This relationship is a direct consequence of the momentum equation and mass conservation law for electrons.  This is the first measurement of the relationship between the density fluctuations and the wave potential although it has been studied
previously {[}Neugebauer, 1975; Kellogg et al, 1999{]}.
\begin{figure}[bt] 
\centering
\includegraphics[width=\linewidth]{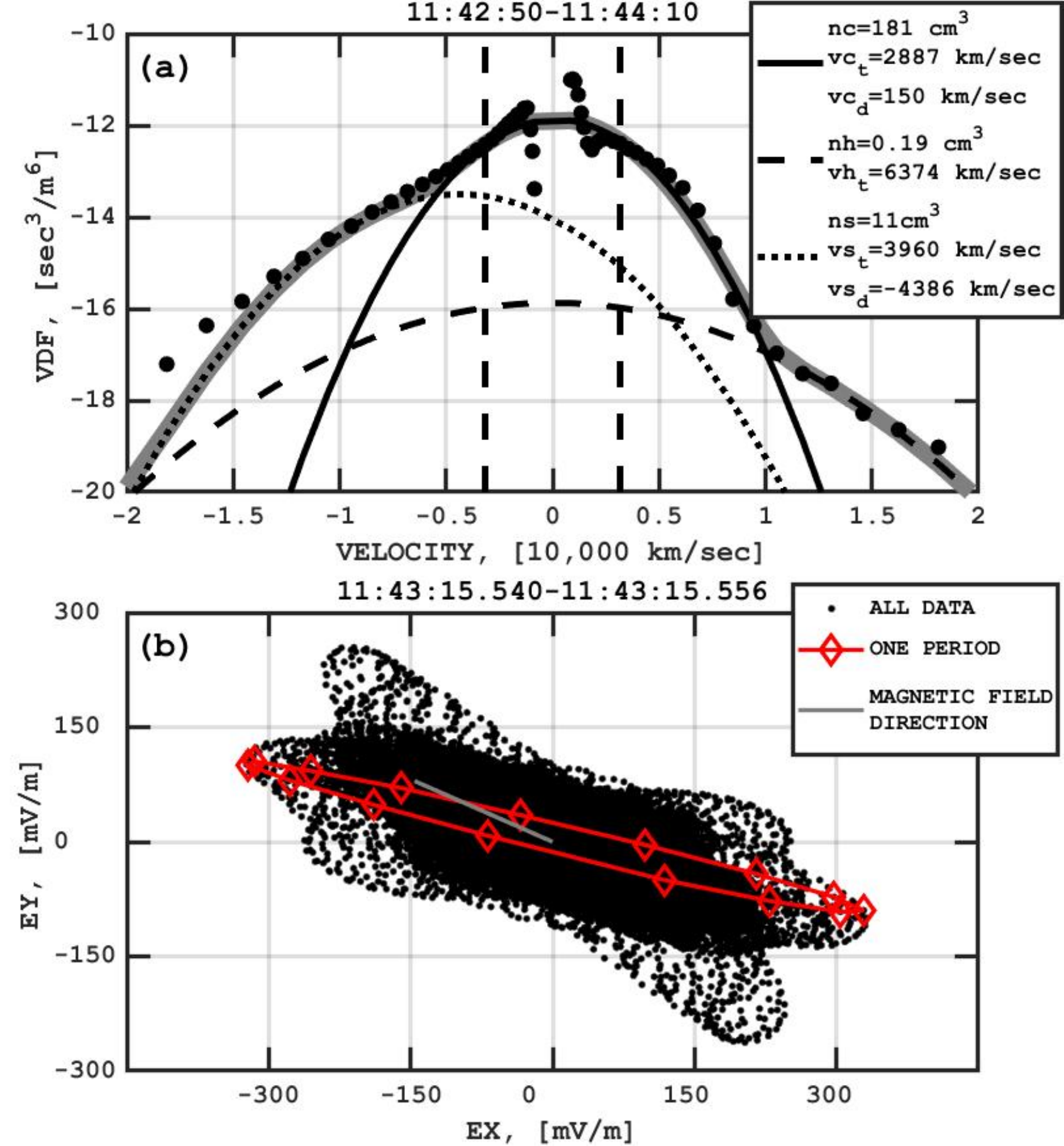}
\caption{
Panel 3a gives the electron velocity distribution function
during an 80 second interval surrounding the 15-millisecond large
Langmuir wave. It describes the core, halo and strahl distributions
which show that the average core electron density and thermal velocity
were 181 cm\textsuperscript{-3} and 2887 km/sec. Panel 3b gives the
hodogram of the electric field during this event. The hodogram is in the
X-Y plane but, because $B_Z\approx0$, it describes
the relationship between the electric field and the magnetic field
direction, which is the green line. It is seen that the electric field
is generally parallel to the magnetic field with significant deviations
that are described in Figure 2d. The red curve in Figure 3b represents
one period of the wave at the time of the data in Figure 5.
}
\end{figure}

Figure 3a shows the electron velocity distribution function (vdf) during
an 80 second interval surrounding the 15-millisecond large field event
of Figure 2, using data obtained from on-board electron measurements
{[}Whittlesey et al, 2020{]}. Because this data collection interval is
about 5000 times greater than the duration of Figure 2 and because
Langmuir waves appeared only a small fraction of this time (see Figure
1a), the vdf of Figure 3a is an average over an interval that is largely
devoid of Langmuir waves. The currents obtained from the core and strahl
distributions are 4.3 and -7.7 microamps/m\textsuperscript{2}.

Figure 3b presents the hodogram of the electric field in the X-Y plane.
Because, as noted, $B_z \approx 0$, the hodogram
illustrates the relationship between the electric field $k$-vector and the
magnetic field (the green line in Figure 3b). From 3b it is seen that
the electric field was generally parallel to the background magnetic
field with deviations that are described in Figure 2d.

Figure 4 presents spectra of the electric field (black), magnetic field
(red) and density fluctuations (green) for the largest event in Figure 2. (Similar results for both a small and a medium amplitude event are presented in the appendix). All three quantities in Figure 4 
had spectral peaks near the plasma frequency of 125 kHz, with the magnetic
field peak previously presumed to result from interaction of the plasma
oscillations with ion sound waves to produce Z-mode waves {[}Gurnett and
Anderson, 1976; Hospodarsky and Gurnett, 1995{]}. At the seven harmonics
of the Langmuir frequency, prominent electric fields were observed
having no measurable magnetic field components. 
If magnetic fields with amplitudes relative to that at the Langmuir frequency existed,
they would be below the noise threshold of the measurement.
That these harmonics are real and
not artifacts of an imperfect measurement, is clear from the fact that,
if they were not real, there would be no harmonics and no type-III
emission.
It is noted that other examples of wave spectra are given in Figures A1 and A2.
\color{black}
\begin{figure}[tb] 
\centering
\includegraphics{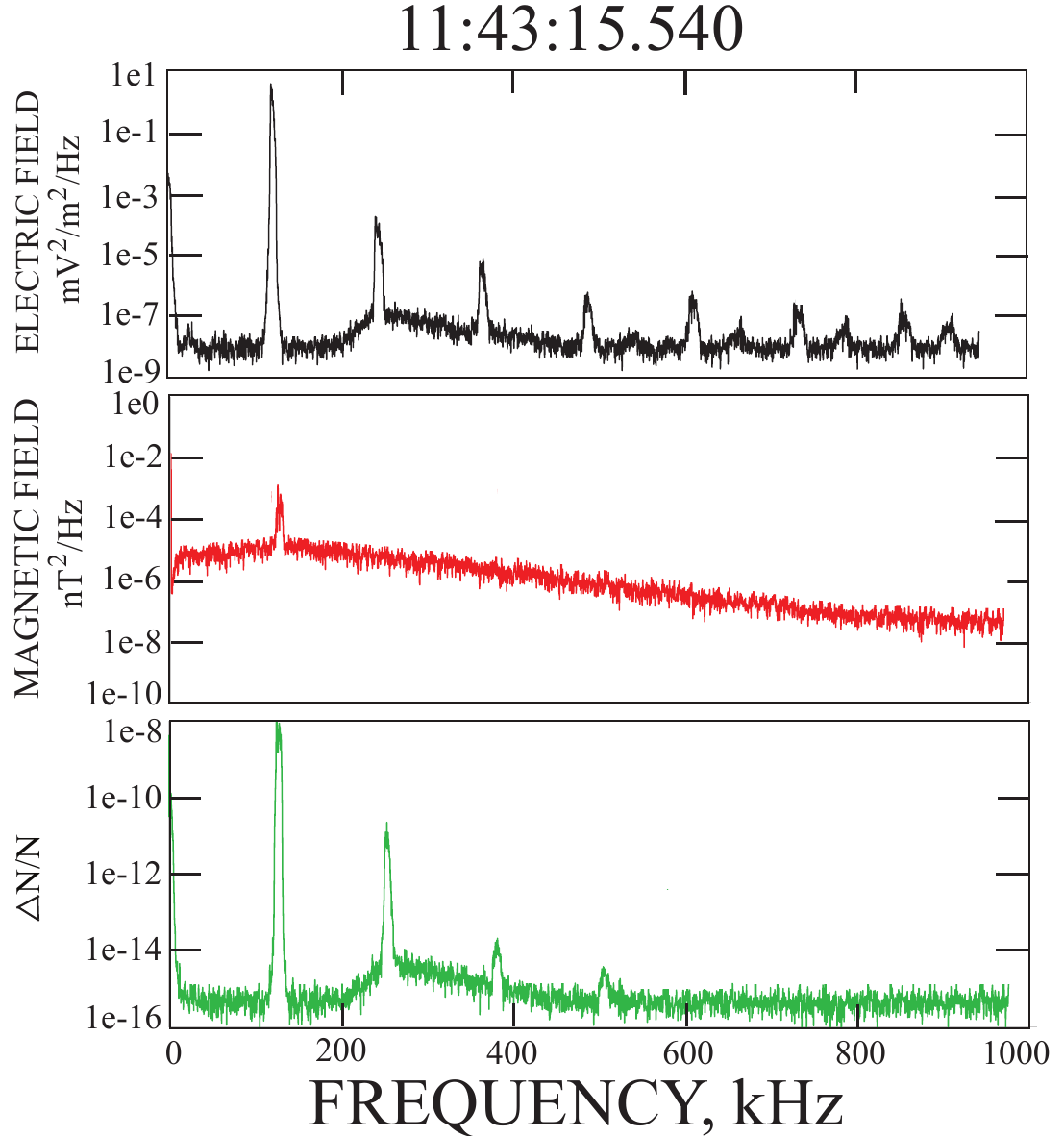}
\caption{
Spectra of the electric field, the magnetic field and the
density fluctuations observed during the intense burst of Figure 2. The
lowest frequency wave, at 125 kHz, is at the Langmuir frequency and it has both a
magnetic and density fluctuation component.
Six electric field harmonics of the intense burst are observed and three
of them have detectable density fluctuations.
}
\end{figure}

That the electric field instrument is capable of measurements at the
frequencies observed in Figure 4 is verified from the fact that the time
required to change the spacecraft potential by 1 volt through charging
its $1\times 10^{-10}$ farad capacitance by the photoemission current of 
about $5 \times 10^{-3}$ amps is about $2\times 10^{-8}$ seconds.

Theories of such harmonic waves, produced in density cavities {[}Ergun
et al, 2008; Malaspina et al, 2012{]}, do not apply to the present
observations because density cavities were not present at times of any
of the events discussed on the day of interest. Instead, the harmonics
were likely created by a single wave in an electron beam that grew
exponentially until the beam electrons were trapped. At that time, the
wave amplitude stopped growing and began to oscillate about a mean
value. During the trapping process, the beam electrons were bunched in
space and higher harmonics of the electric field were produced {[}O'Neil
et al, 1971, 1972{]}. 

\begin{figure}
\centering
\includegraphics[width=\columnwidth]{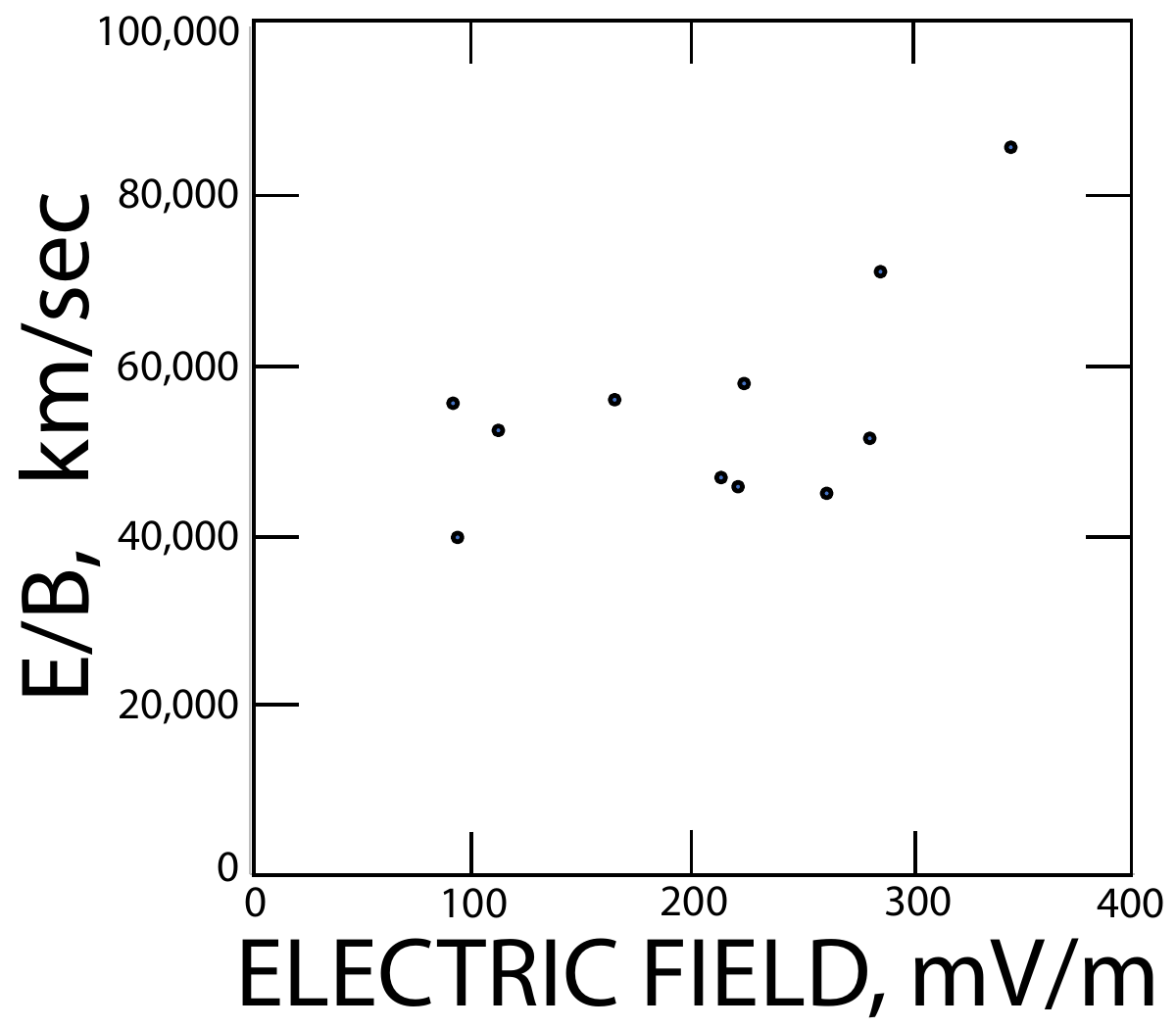}
\caption{
A plot of a single component of the electric field divided by a single component of the magnetic field for the eleven events depicted in Figure 2.  For a full measurement, this ratio is equal to the phase velocity of the wave. The ratio is an overestimate of the phase velocity of the electromagnetic wave because part of the electric field resulted from an electrostatic wave.  Thus, the phase velocity of the electromagnetic wave was <50,000 km/sec. 
}
\end{figure}

Properties of the electromagnetic waves near the Langmuir frequency may be obtained by considering the ratio of their electric fields to their magnetic fields, which, for a full field measurement, gives the phase velocity of the wave.  Although only two components of E and one component of B were measured, their ratio can give an estimate of the phase velocity.  In Figure 5, this ratio is plotted for the 11 events in Figure 2, and the average velocity thereby obtained is about 50,000 km/sec.  This is an overestimate of the velocity because part of the electric field was in the electrostatic, not the electromagnetic, wave. With a frequency of 125 kHz, this wave had a wavelength that was less than 400 meters and a k-value greater than 0.015. From the electron core temperature of 23 eV and the plasma density of 181
cm$^{-3}$, both obtained from the vdf of Figure 3a, the
Debye length was 2.65 meters. Thus, 
$k \lambda_d$ was greater than 0.04.

The phase velocity of the Langmuir wave may be estimated from equation (1), using the measured electric field of 250 mV/m and density fluctuations of 0.02 in Figure 2. With uncertainties in each of these quantities of about a factor of two, the k-value obtained from equation (1) is 0.14. This gives $k \lambda_d$  = 0.4 and a phase velocity of the electrostatic wave that was less than 10,000 km/sec.

When a positive electric field in the wave passes over the spacecraft, $V1$ (which is the potential of antenna 1 minus the spacecraft potential), first becomes positive and then, $V2$ becomes negative (because the spacecraft potential becomes positive before the antenna potential becomes positive).  Their time difference gives the time required for the wave to propagate across the antenna system, which is proportional to the phase velocity of the wave.  While the phase velocity of the electrostatic wave can be estimated in this way, in fact this time difference actually varied over a range so wide that such a measurement was not possible.  In addition, the ratio of the few volt amplitude of $V1$ to the amplitude of $V2$ varied by almost two orders-of-magnitude, as illustrated in Figure 6a.  These facts must be due to the presence of two waves whose amplitudes at each antenna add together to produce the observed wave.  However, two waves with constant and different frequencies add to produce beats like those in Figure 6c, and this is not what is observed.  Because the Langmuir wave frequency is proportional to the square root of the density, its frequency varied with time as it passed over density irregularities.  As seen from the density plot of Figure 6b, this is an important effect because the density varied by as much as 10 percent in the vicinity of the observations.  Allowing the Langmuir wave frequency to vary pseudo-randomly by one or two percent would produce many different curves, a typical one of which is illustrated from a simulation in Figure 6d. Thus, it is concluded that the signal near the Langmuir frequency is made of an electromagnetic wave having a phase velocity of about 25,000 km/sec and a Langmuir wave whose frequency varied as it passed over density irregularities.  As seen in appendix Figure A3, the harmonics of two other waves each also contained two waves, one of which had a frequency that also varied during its passage over the spacecraft because such variations of the voltage ratio and time delay also occurred in that data.

It is important to note that the rapid (less than one millisecond) amplitude variations typical of the Langmuir wave and its harmonics are artifacts resulting from addition of two waves,  and that the observed wave amplitude is not representative of the true wave amplitude versus time.  

\begin{figure}[t] 
\centering
\includegraphics[height=4in]{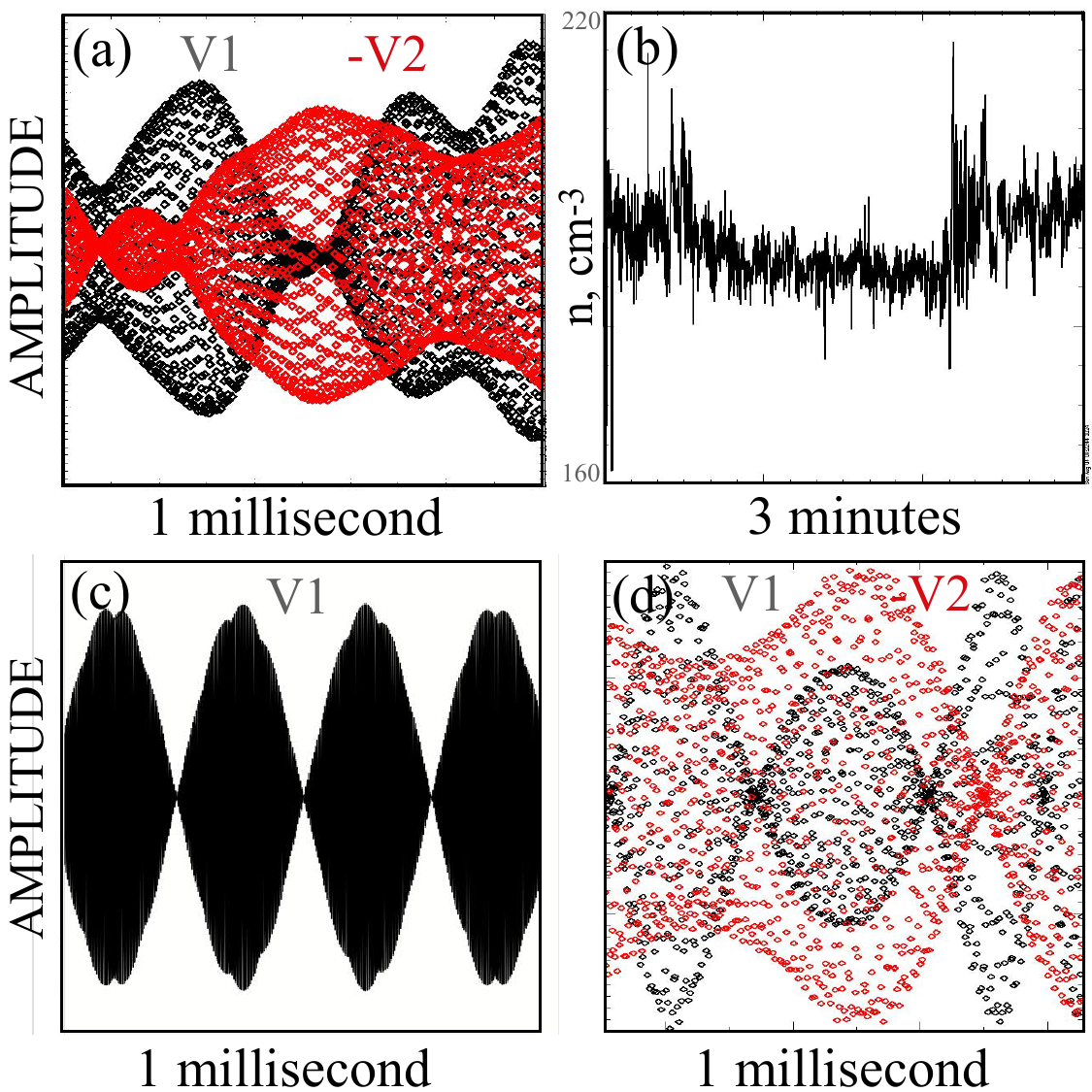}
\caption{
Panel 6a gives the few volt potentials on antennas 1 and 2 during a one millisecond interval.  Their difference must be due to the addition of two waves at each antenna, with the Langmuir wave having a variable frequency to avoid the beats produced by two fixed-frequency waves, as illustrated in panel 6c.  The Langmuir wave frequency varied because of the large density irregularities illustrated in panel 6b, to produce a simulation in panel 6d of the signals on the two antennas that are produced by one or two percent Langmuir wave frequency variations.
}
\end{figure}

\section{Discussion}

The above analyses show that slow electrostatic waves (phase velocity < 10,000 km/sec) existed near the Langmuir frequency and at as many as six harmonics, the number of which increases with the amplitude of the Langmuir wave.  Their electrostatic nature is shown by measurements of density fluctuations of the appropriate magnitudes at each frequency.  These harmonics
were likely created by a single wave in an electron beam that grew
exponentially until the beam electrons were trapped. At that time, the
wave amplitude stopped growing and began to oscillate about a mean
value. During the trapping process, the beam electrons were bunched in
space and higher harmonics of the electric field were produced {[}O'Neil
et al, 1971, 1972{]}. Conversion of these harmonics to electromagnetic waves via propagation through an inhomogeneous plasma having density fluctuations is a process that has been well-studied {[}Stix, 1965; Oyo, 1971; Melrose, 1980a; Melrose, 1981; Mjolhus et al, 1983; Morgan and Gurnett, 1990; Kalaee et al, 2009;
Schleyer et al, 2013; Kim et al, 2013; Kalaee et al, 2016{]}. It is
plausible that the Langmuir frequency harmonics were converted to type-III
emissions in this way.

It is emphasized that none of these results are required of or explained by the conventional model of the three-wave interaction of two counter-streaming Langmuir waves that coalesce to produce the type-III wave.  In fact, no evidence has been found in this or earlier studies that requires this coalescence model for its understanding.

There is also an electromagnetic wave near the Langmuir frequency having a phase velocity less than 50,000 km/sec.  Whether there are electromagnetic waves at the harmonics of the Langmuir frequency cannot be determined because, if they existed, their magnetic field components would be below the noise level of the measurement.  A possible explanation of this wave and the existence of a slow electrostatic wave is next given.

The following discussion of current-driven generation of electrostatic and
electromagnetic fields at frequencies very close to the electron plasma
frequency, $\omega_p$, is based on the results of previous
studies {[}Sauer and Sydora, 2015, 2016; Sauer et al, 2017, 2019{]} on
the excitation of current-driven Langmuir oscillations with
$\omega = \omega_e$ and $k=0$ by which, subsequently, a parametric decay
in Langmuir and ion-acoustic waves with finite wave number is triggered.
It is assumed that the electron component is composed of two parts: a
dominant population at rest and a minor component in relative motion, as
seen in the core, halo and strahl of Figure 3. The mechanism of wave
excitation does not take into account beam instabilities that may occur
in an earlier phase of the interaction. The electron current resulting
from the relative motion between the two electron components is crucial
for triggering the Langmuir oscillations.

\begin{figure}[t] 
\centering
\includegraphics[height=4in]{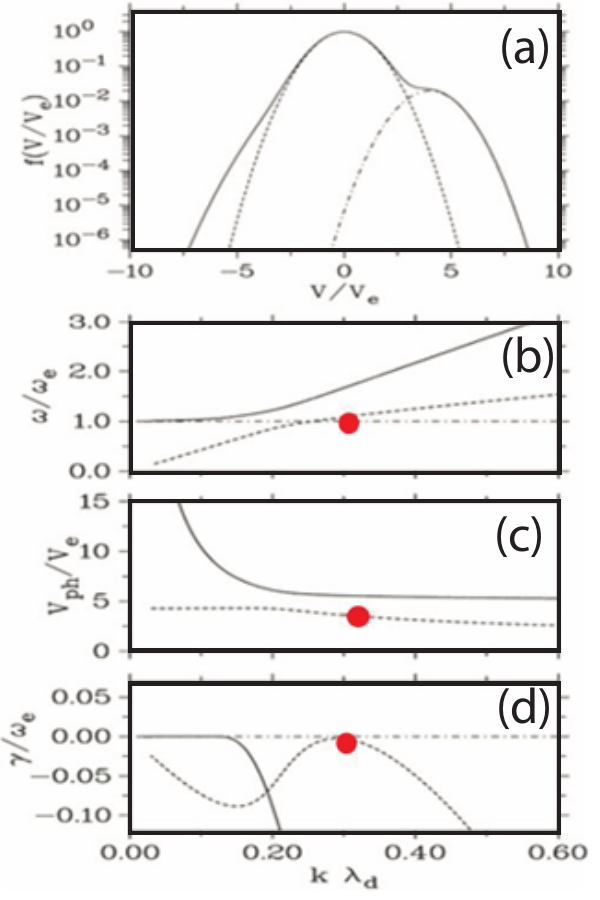}
\caption{
a) Model electron velocity distribution function (EVDF) of the
solar wind consisting of core, halo and strahl and the associated
dispersion of the Langmuir/electron-acoustic mode. b) frequency
$\omega / \omega_e$, c) phase velocity $V_ph / V_e$, where $Ve$ is 
the thermal core velocity, and d) damping rate $\gamma / \omega_e$.
}
\end{figure}

A characteristic electron velocity distribution function of the solar
wind is shown in Figure 7a together with the related dispersion of
Langmuir waves in the wave number range up to $k \lambda_d =0.6$.
As seen in Figure 7b, due to the occurrence of the electron-acoustic
mode which is associated with the strahl, mode splitting takes place at
the point where the Langmuir mode of the main population is crossed,
that is at $k \lambda_d \approx 0.3$. Correspondingly, the
related phase velocity is
$V_{ph} \approx 3.5 V_e$
(Figure 6b).
Around this point the Langmuir/electron-acoustic mode is weakly damped
(Figure 7c) and its frequency approaches the plasma frequency,
$\omega_p$, (see red points). As shown in earlier papers
{[}Sauer and Sydora, 2015, 2016, Sauer et al, 2017, 2019{]}, the current
due to the strahl drives Langmuir oscillations at $\omega_e$ and
simultaneously Langmuir/electron-acoustic waves are driven by parametric
decay. Their wave length is given by that of the electron-acoustic mode
at $\omega \approx \omega_e$,
which in the present case is
$k \lambda_d \approx 0.3$.

Four arguments supporting the validity of the observed density data are:

\begin{enumerate}
\def\labelenumi{\arabic{enumi}.}
\item
  In other publications, density fluctuations obtained from the
  spacecraft potential have been observed and shown to be valid at low
  and high frequencies {[}Roberts et al, 2020; Mozer et al, 2023{]}.
  Such fluctuations have also been observed in laboratory measurements
  {[}Hu, 2021{]}.
\item
  Figure 1c illustrates the Langmuir frequency obtained from the plasma
  density which, in turn, was obtained from the spacecraft potential.
  That the plasma frequency of Figure 1c agrees with the measured plasma
  frequency of Figure 1a is proof of the validity of the low frequency
  density estimate obtained from the spacecraft potential.
\item
  Equation (1) presents the theoretical relationship between the
  electric potential (or field) in the wave and the density
  fluctuations, $\Delta N / N$, that is required in an electrostatic wave. That
  the ratio of the electric field to $\Delta N / N$ in this equation is within a
  factor of two equal to the more variable measured ratios, is evidence
  that determination of the density from the spacecraft potential is
  valid at high frequency.
\item
  A further requirement of equation (1) is that the electric field and
  density fluctuation in the high frequency wave be 90 degrees out of
  phase. This has been found to be the case (see Figure A4 in the appendix) which provides   
  conclusive evidence that the electric field and $\Delta N / N$.
  were well-measured 
  and that they were not artifacts of a poor measurement.
\end{enumerate}     

Previously observed Langmuir wave electric fields have ranged from a few
to about 100 mV/m {[}Gurnett and Anderson, 1976; Graham and Cairns,
2012; Sauer et al, 2017{]}. This range is associated with the range of
driving current densities in the plasma. The observed electric field of
about 300 mV/m is a factor of three larger than the largest of the
previously observed Langmuir waves {[}Malaspina et al 2010{]}. The
explanation of this difference may lay in the fact that the Langmuir waves
had a wavelength that was longer than the
three-meter Parker Solar Probe antenna but was similar to or shorter than the
typical antennas that made the earlier measurements. An antenna that is
longer than the wavelength of the wave, under-estimates the amplitude of
the true electric field because of the many wavelengths of the field
that are inside the antenna. This antenna property may explain why the
largest earlier electric field measurements were smaller than observed
in the current example.

The rapid (less than one millisecond) amplitude variations typical of the Langmuir wave and its harmonics are artifacts resulting from addition of two waves, one of which has small frequency variations that arise from traveling through density irregularities.

\section{Acknowledgements}

This work was supported by NASA contract NNN06AA01C. The authors
acknowledge the extraordinary contributions of the Parker Solar Probe
spacecraft engineering team at the Applied Physics Laboratory at Johns
Hopkins University. The FIELDS experiment on the Parker Solar Probe was
designed and developed under NASA contract NNN06AA01C. Our sincere
thanks to J.W. Bonnell, M. Moncuquet, and P. Harvey for providing data
analysis material and for managing the spacecraft commanding and data
processing, which have become heavy loads thanks to the complexity of
the instruments and the orbit.

\section{References}

Bale, S. D., Burgess, D., Kellogg, P. J., Goetz, K., Howard, R. L., \&
Monson, S. J., 1996. GRL, 23, 109--112.
\url{https://doi.org/10.1029/95GL03595}

Bale, S.D., Kellogg, P.J., Goetz, K., and Monson, S.J., 1998, GRL, 25,
NO. 1, 9-12

Bale, S. D., Goetz, K., Harvey, P. R., et al. 2016, SSRv, 204, 49

Bellan, P.M. 2006, Cambridge University Press, https://doi.org/10.1017/CBO9780511807183

Dulk, G. A. 1985, ARA\&A, 23, 169

Ergun, R.E., Malaspina, D.M., Cairns, I,H, et al, 2008, \emph{PRL}
\textbf{101}

Ginzburg, V.L. and Zheleznyakov, V.V., Sov. Astron. AJ 2, 653 (1958)

Goldman, M. V., Reiter, G.F. and Nicholson, D.R., 1980, Phys. Fluids,
23, 388 -- 401.

Graham, B., and Cairns, I.H., 2012, JGR: SPACE PHYSICS, 118, 3968--3984,
doi:10.1002/jgra.50402

Gurnett, D. A. and Anderson, R. R. 1976.~Science.~194~(4270): 1159--1162

Henri, P., Briand, C., Mangeney, A., Bale, S. D., Califano, F., Goetz,
K., \& Kaiser, M., 2009, JGR, 114(A13), 3103.
https://doi.org/10.1029/2008JA013738

Hospodarsky, G.B. and Gurnett, D.A., 1995, GRL, \textbf{22}, NO. 10,
1161-1164

Hu, Y., Yoo, J., Ji, H., Goodman, A., and Wu, X. ,2021, \emph{Rev. Sci.
Instrum.} \textbf{92}, 033534; doi: 10.1063/5.0035135

Jebaraj,I.C., Krasnosalskikh, V., Pulupa, M., Magdalenic, J., and Bale,
S.D. , 2023, Ap.J\textbf{.,} \textbf{955:L20}
\url{https://doi.org/10.3847/2041-8213/acf85}

Kalaee, M.J., Ono, T. Katoh, Y, Lizima, M., and Nishimura, Y., 2009,
Earth Planets Space, \textbf{61}, 1243--1254

Kalaee, M.J., Katoh, Y., 2016, \emph{Phys. Plasmas,}~\textbf{23},
072119, \url{https://doi.org/10.1063/1.4958945}

Kellogg, P.J., 1980, Ap.J., 236, 696-700

Kellogg, P.J., Goetz, K, and Monson, S.J., 1999, JGR, 104, A8,
17,069-17,078

Kim, E.H., Cairns, I.H. and Johnson, J.R., 2013, \emph{Phys. Plasmas},
\textbf{20}, 122103

Larosa, A., Dudok de Wit, T., Krasnoselskikh1, V., Bale, S.D., Agapitov,
O., Bonnell, J.W., et al, 2022, Ap.J., 927:95 (8pp)
\url{https://doi.org/10.3847/1538-4357/ac4e85}

Malaspina, D.M., Cairns, I.H., and Ergun, R.E. 2010, JGR, 115, A01101,
doi:10.1029/2009JA014609

Malaspina, D.M., Cairns, I.H., and Ergun, R.E. 2012, Ap.J.,
\textbf{755:45,} doi:10.1088/0004-637X/755/1/45

Malaspina, D. M., Graham, D.B., Ergun, R.E.,and Cairns, I.H., 2013,

JGR, Space Physics, \textbf{118}, 6880--6888, doi:10.1002/2013JA019309

Melrose, D. B. 1980, Space Sci. Rev., 26, 3

Melrose, D.B., 1980a, Aust. J. Phys, 33,661

Melrose, D.B., 1981, Journal of Geophysical Research: Space Physics,
\textbf{86}, Issue A1 / p. 30-36\textbf{,} https:/
/doi.org/10.1029/JA086iA01p00030

Melrose, D. B. 1987, Sol. Phys., 111, 89

Mjolhus, E., 1983, Journal of Plasma Phy, \textbf{30} Issue 2 pp. 179 -
192 DOI: \url{https://doi.org/10.1017/S0022377800001}

Morgan, D.D., and Gurnett, D.A., 1990, \emph{Radio
Sci.,}\textbf{25}(6),~1321--1339,
doi:\url{https://doi.org/10.1029/RS025i006p01321}

\href{https://ui.adsabs.harvard.edu/search/q=author:\%22Mozer\%2C+F.+S.\%22\&sort=date\%20desc,\%20bibcode\%20desc}{Mozer,
F.S.,~}\href{https://ui.adsabs.harvard.edu/search/q=author:\%22Bale\%2C+S.+D.\%22\&sort=date\%20desc,\%20bibcode\%20desc}{Bale,
S.D.,~}\href{https://ui.adsabs.harvard.edu/search/q=author:\%22Kellogg\%2C+P.+J.\%22\&sort=date\%20desc,\%20bibcode\%20desc}{Kellogg,
P.J.,~}\href{https://ui.adsabs.harvard.edu/search/q=author:\%22Larson\%2C+D.\%22\&sort=date\%20desc,\%20bibcode\%20desc}{Larson,
D.,~}\href{https://ui.adsabs.harvard.edu/search/q=author:\%22Livi\%2C+R.\%22\&sort=date\%20desc,\%20bibcode\%20desc}{Livi,
R.,~}\href{https://ui.adsabs.harvard.edu/search/q=author:\%22Romeo\%2C+O.\%22\&sort=date\%20desc,\%20bibcode\%20desc}{Romeo,
O., 2022,~}The Astrophysical Journal, 926, Issue 2,
DOI:\href{https://ui.adsabs.harvard.edu/link_gateway/2022ApJ...926..220M/doi:10.3847/1538-4357/ac4f42}{10.3847/1538-4357/ac4f42}

Mozer, F.S., Agapitov, O., Bale, S.D., Livi, R., Romeo, O., Sauer, K.,
Vasko, I.Y., Verniero, J., 2023,
\emph{ApJL,}~\textbf{957}~\textbf{DOI}~10.3847/2041-8213/ad0721

Neugebauer, M., 1976, Geophys. Res., 81, 2447, 19

\href{https://ui.adsabs.harvard.edu/search/q=author:\%22O'Neil\%2C+T.+M.\%22\&sort=date\%20desc,\%20bibcode\%20desc}{O'Neil,
T.
M.,~}\href{https://ui.adsabs.harvard.edu/search/q=author:\%22Winfrey\%2C+J.+H.\%22\&sort=date\%20desc,\%20bibcode\%20desc}{Winfrey,
J.
H.,~}\href{https://ui.adsabs.harvard.edu/search/q=author:\%22Malmberg\%2C+J.+H.\%22\&sort=date\%20desc,\%20bibcode\%20desc}{Malmberg,
J. H.} ,1971,\emph{Physics of Fluids}, \textbf{14}, 6, p.1204-1212
\textbf{DOI:}\href{https://ui.adsabs.harvard.edu/link_gateway/1971PhFl...14.1204O/doi:10.1063/1.1693587}{10.1063/1.1693587}~

\href{https://ui.adsabs.harvard.edu/search/q=author:\%22O'Neil\%2C+T.+M.\%22\&sort=date\%20desc,\%20bibcode\%20desc}{O'Neil,
T.
M.,~}\href{https://ui.adsabs.harvard.edu/search/q=author:\%22Winfrey\%2C+J.+H.\%22\&sort=date\%20desc,\%20bibcode\%20desc}{Winfrey,
J. H.,~}1972, \emph{Phys. Fluids}~15, 1514--1522 (1972)
\url{https://doi.org/10.1063/1.1694117}

Oyo, H., 1971, Radio Science, \textbf{6}, Issue 12, pp. 1131-1141
DOI:\href{https://ui.adsabs.harvard.edu/link_gateway/1971RaSc....6.1131O/doi:10.1029/RS006i012p01131}{10.1029/RS006i012p01131}

Papadopoulos, K., Freund, H.P. 1978, Geophys. Res. Lett., 5, 881 -- 884

Pedersen, A., Cattell, C., Falthammar, C-G., Formisano, V., Lindqvist, P-A., Mozer, F., and Torbert, R., 1984, Space Sci. Rev. 37, 269-312

Reiner, M.J. and MacDowall, R.J., 2019, Solar Phys (2019) 294:91
\url{https://doi.org/10.1007/s11207-019-1476-9}

Roberts, O.W., Nakamura, R., Torkar, K., Graham, D.B., Gershman, D.J.,
Holmes, J.C., et al. 2020, \emph{J.G.R., Space Physics}, \textbf{125},
https://doi.org/ 10.1029/2020JA02785

Sauer, K., and Sydora, R., 2015, JGR, Space Physics, 120,
doi:10.1002/2014JA020409.

Sauer, K., and Sydora, R., 2016, Geophys. Res. Lett., 43,
doi:10.1002/2016GL069872.

Sauer, K., Malaspina, D.M., Pulupa, M. and Salem, C.S., 2017, J.
Geophys. Res. Space Physics, 122, 7005--7020, doi:10.1002/2017JA024258

Sauer, K., Baumgärtel, K., Sydora, R., and Winterhalter, D. 2019, JGR,
Space Physics, 124, 68--89. \url{https://doi.org/10.1029/2018JA025887}

\href{javascript:;}{Schleyer}; F., Cairns, I.H. and Kim, E.-H. 2013,
\emph{Phys. Plasmas}~\textbf{20}, 032101,
\url{https://doi.org/10.1063/1.4793726}

Smith, D. F. 1970, Sol. Phys., 15, 202

Smith, R. A., Goldstein, M. L., and Papadopoulos, K. 1976, Sol. Phys.,
46, 515

Stix, T.H. ,1965, \emph{Phys. Rev. Lett.}, \textbf{15}, 23

Sturrock, P. A. 1964, NASA Special Publication, 50, 357

Whittlesey, P.L., Berthomier, M., Larson, D.E., Case, A.W., Kasper,
J.C., et al, 2020, Ap.J., 246, https://doi.org/10.3847/1538-4365/ab7370

Zheleznyakov, V. V., \& Zaitsev, V. V. 1970, Soviet Ast., 14, 47~

%
%

\begin{appendix} 
\renewcommand{\thepage}{A\arabic{page}} 
\renewcommand{\thesection}{A\arabic{section}}  
\renewcommand{\thetable}{A\arabic{table}}  
\renewcommand{\thefigure}{A\arabic{figure}}

\section{Power spectra of a 30 mV/m Langmuir wave}
\begin{figure}[h]
\centering
\includegraphics[width=\linewidth,clip]{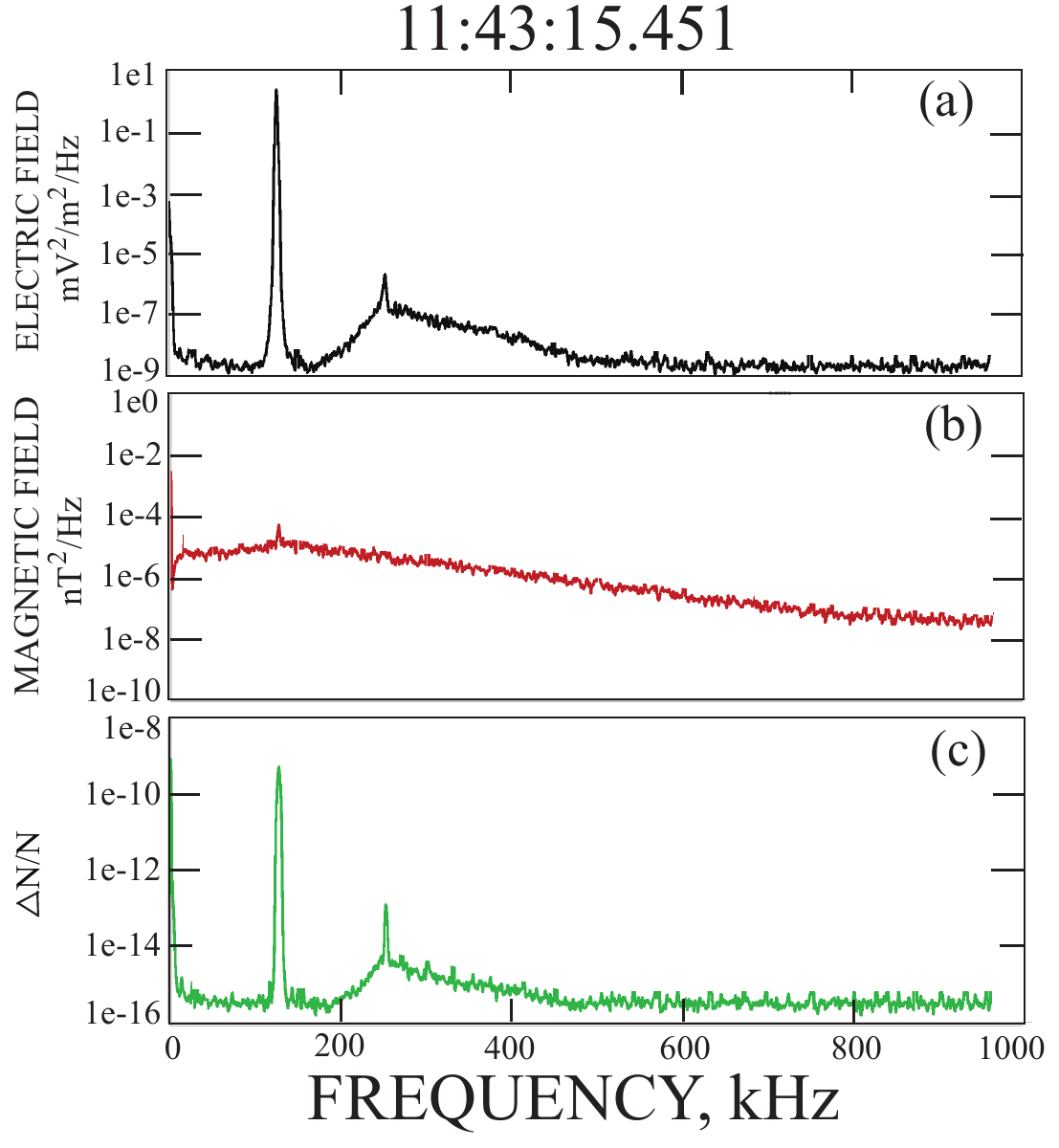}
\caption{Spectra of a 30 mV/m Langmuir wave.}
\label{fig:A1}
\end{figure}

In Figure \ref{fig:A1}, the spectra associated with a lower amplitude Langmuir wave are displayed.  Although this wave was an order-of-magnitude smaller than the wave having the spectra of Figure 4, the peaks near the plasma frequency and its first harmonic of the electric field of panel A1(a), the magnetic field of panel A1(b) and the density fluctuations of panel A1(c) show that this wave and its harmonic were also slow electrostatic waves.
\FloatBarrier
\newpage
\section{Power spectra of a 150 mV/m Langmuir wave}
\begin{figure}[h]
\vspace{.45in}
\centering
\includegraphics[width=\linewidth,clip]{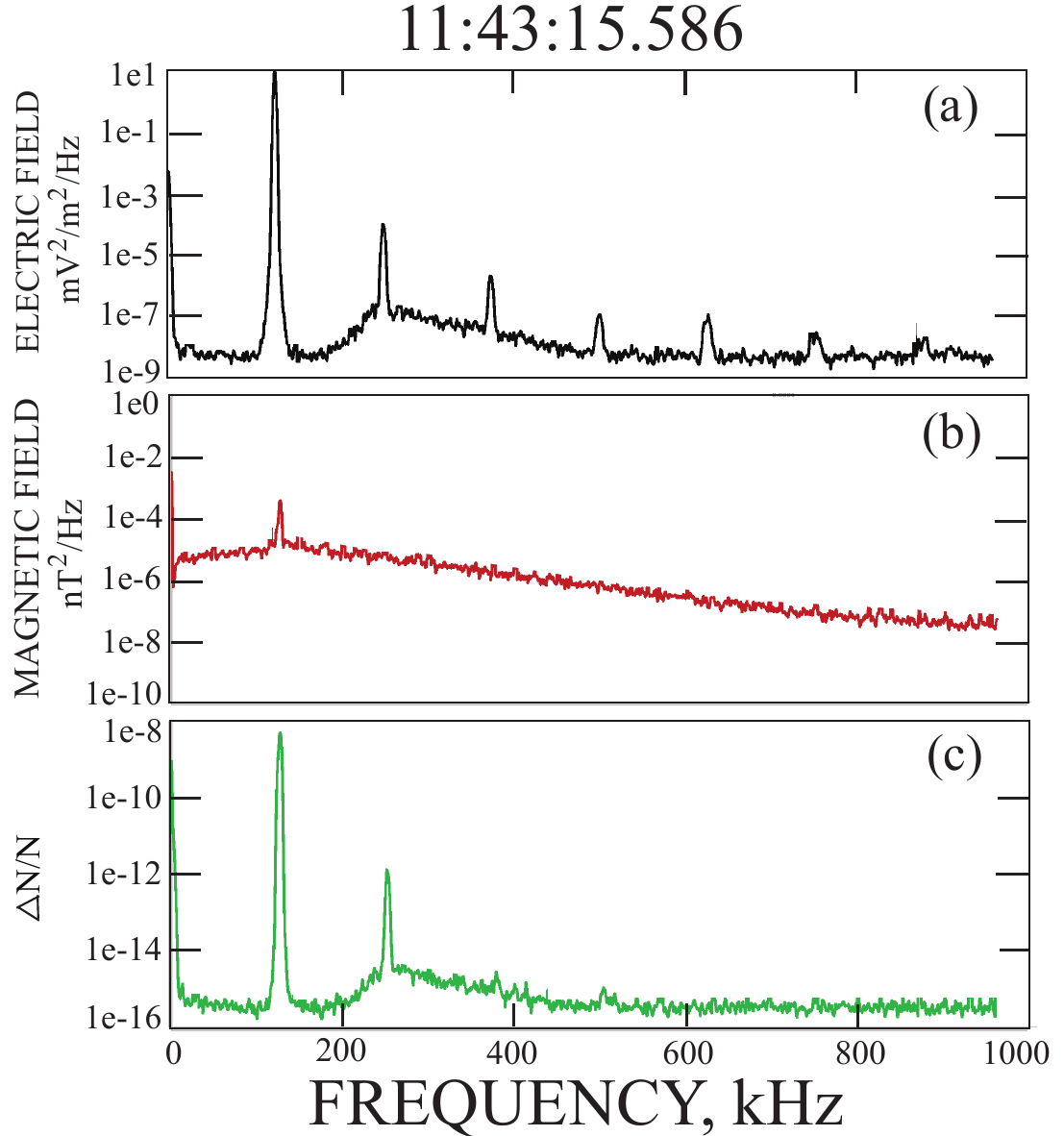}
\caption{Spectra of a 150 mV/m Langmuir wave}
\label{fig:A2}
\end{figure}

The wave of Figure \ref{fig:A2} also shows spectral peaks similar to those observed for a 300 mV/m wave (Figure 4) and a 30 mV/m wave (Figure A1).  These same peaks were observed for all waves in Figure 2, so they were typical of the waves observed during the type-III emission.. 
\FloatBarrier
\newpage
\section{Antenna potentials for plasma frequency and first harmonic waves observed at two different times}
\begin{figure}[h]
\centering
\includegraphics[width=\linewidth,clip]{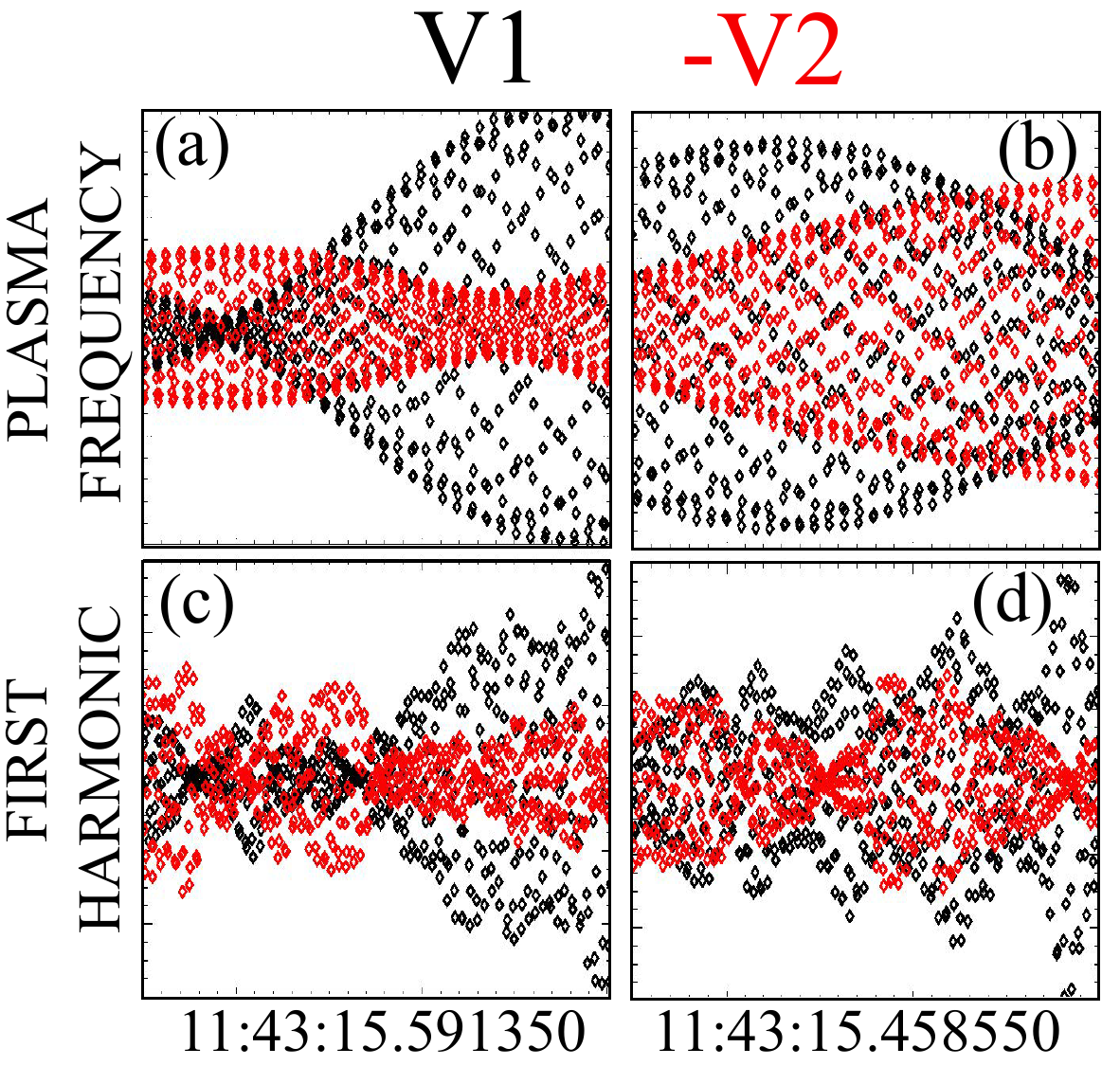}
\caption{Single-ended antenna potentials for two wave events.}
\label{fig:A3}
\end{figure}

Figure \ref{fig:A3} illustrates the single-ended potentials $V1$ and $-V2$ for the plasma waves of panels A3(a) and A3(b) and their first harmonics in panels A3(c) and A3(d) at the time of the low amplitude plasma waves in panels A3(b) and A3(d) and the medium amplitude plasma waves in panels A3(a) and A3(c).  Because, in all cases, the ratio of the two potentials varied by large factors, pairs of electrostatic waves existed during both the events with at least one of the pair being slow.  This same behavior was observed for all the bursts of Figure 2.
\FloatBarrier
\newpage
\section{Relative timing of the electric field and the density fluctuations}
\begin{figure}[h]
\centering
\includegraphics[width=\linewidth,clip]{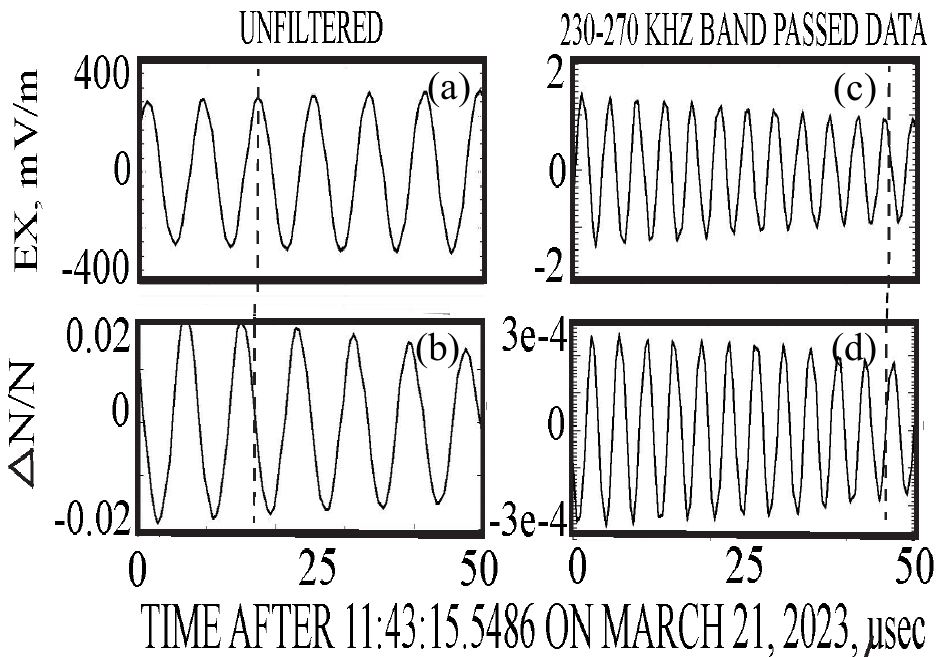}
\caption{Relative timing of the electric field and density waves.}
\label{fig:A4}
\end{figure}

Equation 1 requires that the electric field and density components of a wave be 90 degrees out of phase.  This requirement is verified in Figure \ref{fig:A4}. which illustrates several cycles of the electric field and density fluctuations near the plasma frequency in panels A4(a) and A4(b) and the first harmonic in panels A4(c) and A4(d).  For each case a vertical dashed line allows comparison of the phase difference between the E-field and the density wave, which is found to be close to 90 degrees.  This result supports the conclusion that the waves were accurately measured.
\FloatBarrier

\end{appendix}
\end{document}